%% file: Ferdman_0737A_arXiv.tex
\documentclass[letter]{emulateapj}
\usepackage{natbib}
\usepackage{epsfig}

\usepackage{amssymb}
\usepackage{amsmath}
\usepackage{latexsym}
\usepackage{ifthen}
\usepackage{url}
\usepackage{comment}

\newcommand{\degrees}{\,^\circ}
\newcommand{\msun}{\,M_{\odot}}

\newcommand{\kms}{\,\mathrm{km}\,\mathrm{s}^{-1}}

\newcommand{\psr}[1]
{\ifthenelse{\equal{#1}{0737}}{PSR~J0737$-$3039}{\ifthenelse{\equal{#1}{1534}}{PSR~B1534+12}{\ifthenelse{\equal{#1}{1913}}{PSR~B1913+16}{\ifthenelse{\equal{#1}{1802}}{PSR~J1802$-$2124}{\ifthenelse{\equal{#1}{1756}}{PSR~J1756$-$2251}{\ifthenelse{\equal{#1}{1906}}{PSR~J1906+0746}{\ifthenelse{\equal{#1}{1141}}{PSR~J1141$-$6545}{\ifthenelse{\equal{#1}{2127}}{PSR~B2127+11C}{\ifthenelse{\equal{#1}{0751}}{PSR~J0751+1807}{\ifthenelse{\equal{#1}{1713}}{PSR~J1713+0747}{\bf ???????}}}}}}}}}}}

\slugcomment{Draft \today; accepted for publication in the Astrophysical Journal}
\shorttitle{Formation of the double pulsar}
\shortauthors{Ferdman et al.}

\begin{document}

\title{The double pulsar: evidence for neutron star formation without an iron core-collapse supernova}

\author{
R.~D.~Ferdman\altaffilmark{1,2}, 
I.~H.~Stairs\altaffilmark{3}, 
M.~Kramer\altaffilmark{4,1}, 
R.~P.~Breton\altaffilmark{5}, 
M.~A.~McLaughlin\altaffilmark{6}, 
P.~C.~C.~Freire\altaffilmark{4},
A.~Possenti\altaffilmark{7},
B.~W.~Stappers\altaffilmark{1},
V.~M.~Kaspi\altaffilmark{2},
R.~N.~Manchester\altaffilmark{8}, 
A.~G.~Lyne\altaffilmark{1}
}

\altaffiltext{1}{School of Physics and Astronomy, University of Manchester, Jodrell Bank Centre for Astrophysics, Alan Turing Building, Oxford Road, Manchester M13 9PL, United Kingdom; ferdman@jb.man.ac.uk}
\altaffiltext{2}{Department of Physics, McGill University, Rutherford Physics Building, 3600 University Street, Montreal, QC, H3A 2T8, Canada}
\altaffiltext{3}{Department of Physics and Astronomy, University of British 
  Columbia, Vancouver, BC, V6T 1Z1, Canada}
\altaffiltext{4}{Max-Planck-Institut f\"{u}r Radioastronomie, Auf dem H\"{u}gel 69, 53121, Bonn, Germany}
\altaffiltext{5}{School of Physics and Astronomy, University of Southampton, Highfield, Southampton SO17 1BJ, United Kingdom}
\altaffiltext{6}{Department of Physics, West Virginia University, Morgantown, WV 26505} 
%\altaffiltext{7}{National Radio Astronomy Observatory, Green Bank, WV 24944} 
\altaffiltext{7}{INAF - Osservatorio Astronomico di Cagliari, Loc.~Poggio dei Pini, 09012 Capoterra (CA), Italy}
\altaffiltext{8}{CSIRO Astronomy and Space Science, Australia Telescope National Facility, Epping NSW 1710, Australia}

\begin{abstract}
The double pulsar system PSR J0737$-$3039A/B is a double neutron star binary, with a $2.4\,$hr orbital period, which has allowed measurement of relativistic orbital perturbations to high precision. The low mass of the second-formed neutron star, as well as the low system eccentricity and proper motion, point to a different evolutionary scenario compared to most other known double neutron star systems. We describe analysis of the pulse profile shape over 6 years of observations, and present the resulting constraints on the system geometry. We find the recycled pulsar in this system, PSR~J0737$-$3039A, to be a near-orthogonal rotator, with an average separation between its spin and magnetic axes of $90\pm11\pm5\degrees$.
% (the first and second uncertainty values are statistical and systematic, respectively).  
Furthermore, we find a mean $95\%$ upper limit on the misalignment between its spin and orbital angular momentum axes of $3.2\degrees$, assuming that the observed emission comes from both magnetic poles. This tight constraint lends credence to the idea that the supernova that formed the second pulsar was relatively symmetric, possibly involving electron-capture onto an O-Ne-Mg core.  
\end{abstract}


\section{Introduction}
Pulsars are the rapidly rotating neutron star (NS) remnants of supernova (SN) explosions, which emit beamed and highly coherent radiation, most probably from field lines emanating from the magnetic polar regions.
This radiation can be viewed from Earth as a radio pulse each time the beam sweeps across our line of sight.
This pulse represents the variation in flux across the portion of the emission beam that we observe.

In general, NSs that are members of binary systems represent a separate population from the ``standard'' isolated pulsars, which generally have $\sim 1\,$s rotation periods.  Most pulsars observed in binary systems are ``recycled'' to have faster rotation speeds in a phase of mass and angular-momentum transfer from the companion star. 
Many neutron star-white dwarf (NS-WD) systems, and especially double neutron star (DNS) systems, are in close orbits, often displaying relativistic effects.  Measuring the pulse times-of-arrival from these systems has provided some of the most stringent and varied tests of general relativity (GR) available, as well as the most precise measurements of compact object masses.

The double pulsar system, \psr{0737}A/B, was discovered as the most relativistic DNS binary yet to be found \citep{bdp+03} and, as its nickname suggests, the only known system in which both NSs have been visible as pulsars \citep{lbk+04}:
one is a recycled pulsar, \psr{0737}A (henceforth referred to as pulsar ``A''), with a spin period of 22.7 ms; the other pulsar, \psr{0737}B (henceforth referred to as pulsar ``B''), which has a $2.77$-s rotation period, was formed in the second SN and has not been spun up.
The highly relativistic nature of this system, and the presence of two pulsars, has provided pulsar astronomers and NS theorists with many astrophysical phenomena to study in greater detail, and with more precision, than ever before: it immediately provided the best prediction of the coalescence rate of DNS systems \citep{bdp+03, kkl+04};  it also allowed for the measurement of the spin-orbit precession of pulsar B by modeling the eclipse light curves as pulsar A passed behind pulsar B, consistent with the predictions of GR \citep{bkk+08}.  A more recent study of pulsar B has constrained its pulse profile evolution due to geodetic precession, thus constraining models of its emission beam shape, and has confirmed that, due to precession, our line of sight currently intersects an empty portion of the beam \citep{pmk+10}.
In contrast, \citet{mkp+05} found no evidence for secular change in the pulse profile of pulsar A over a 3 year data span, suggesting either a close alignment of the pulsar spin axis with the orbital angular momentum, or a fortuitous precession phase.

Perhaps the best-known result from observations of the double pulsar is the long-term timing analysis that has led to the confirmation of the predictions of GR in the strong-field regime to within $0.05\%$, based on the measurement of Shapiro delay in this system \citep{ksm+06}.  
The parameters describing the \psr{0737}A/B system continue to be monitored and updated, providing increasingly rigid constraints on relativistic theory.  An extensive review and speculations on future science with the double pulsar is given by \citet{ks08}.

\subsection{A Different Evolution?}

\input{dns_table_arXiv}

Several of the properties of the \psr{0737}A/B system differ significantly from other known DNS binaries.  In particular, the low pulsar-B mass of $1.2489(7)\msun$, the small measured eccentricity (0.0878), and low transverse velocity ($10\kms$) \citep{ksm+06}, show a marked departure from some DNS systems for which these parameters have been measured, though also similarities with others, as discussed below.   
This is certainly due to the specific evolution of these systems.  
It is generally thought that DNS systems result from one of three evolutionary scenarios, which we briefly describe here.  A more detailed overview of DNS and binary NS evolution can be found in, e.g., \citet{bv91}, \citet{pk94}, \citet{sta04a}, and \citet{tv06corr}.  

In the \textit{``standard''} scenario, the system initially consists of two massive stars, typically $>8-9\msun$.  The more massive of these eventually evolves off the main sequence, at which time it fills its Roche lobe, stably transferring matter to the secondary star, before collapsing to explode in a SN, leaving behind the first-formed NS in a high-mass X-ray binary \citep[HMXB; see, e.g.,][]{tv06corr}.  The remaining massive secondary then evolves to overfill its Roche lobe.  Unstable mass transfer then occurs, creating a common envelope (CE) that engulfs the NS,  which then spirals in toward the companion, picking up angular momentum and spinning up as a result.  Dynamical friction results in the ejection of the CE, leaving behind a NS with the remaining He core of the companion star.  Further mass transfer may ensue, in which case there is a further increase in the NS rotation speed.  The He star then eventually undergoes a SN, leaving behind the DNS.

The \textit{double-helium core} scenario was hypothesized to avoid possible hypercritical accretion onto the NS during the CE phase that may cause collapse into a black hole, rendering the system unobservable \citep{bro95}.  Here, the two massive progenitor stars initially have masses within $\sim10\%$ of each other, and the orbit is sufficiently wide for the primary to evolve a CO core.  The secondary then evolves off the main sequence before the primary undergoes a SN.  The mass transfer rate in this case is very high, such that the CE that develops is comprised of both stars' envelopes.  This causes a rapid spiral-in of the cores and CE ejection.  After the primary undergoes SN, the remaining evolution resulting in the DNS is similar to the standard process described above.  Although it is expected to occur $2-10$ times less frequently than in the standard picture \citep{dps06}, it may be the predominant formation channel; this 
depends on the likelihood of the standard scenario leading to black hole formation due to hypercritical accretion by the first-formed NS in a CE, thus inhibiting DNS formation \citep{che93}.  On the other hand, it has been shown in the case of the NS-CO white dwarf binary \psr{1802} that it is possible for a close NS binary system to survive a CE phase without hypercritical accretion occurring \citep[][and references therein]{fsk+10}.

A third scenario, commonly referred to as an \textit{electron-capture supernova} (ECS), separates itself in how the secondary star evolves to form a NS.  Here, an O-Ne-Mg core passes a threshold density that allows electrons to be captured on $^{24}$Mg. This decreases the electron degeneracy pressure, which in turn lowers the Chandrasekhar mass of the core, inducing its collapse \citep{mnys80,nom84,pdl+05}.  This would result in low NS velocities, since 
this type of event is thought to proceed over a much shorter timescale than that needed to develop the instabilities thought to cause substantial SN kicks \citep{plp+04, vdh04}, and has also been invoked to explain the formation of a subset of NSs with significantly lower measured masses relative to the overall known NS mass distribution \citep[including PSR~J0737$-$3039B, discussed in this work;][]{spr10}. The ECS scenario is also a suspected mechanism for formation of wide-orbit HMXBs consisting of Be-star donors \citep{lskb09}.
  
In order to investigate how these scenarios may apply to DNS systems we observe, we list
those with measured masses in Table~\ref{tab:dns}, and the values of several of their properties.  PSR~B2127+11C 
resides in the globular cluster M15 \citep{jcj+06}; the evolution of this pulsar thus likely involved unique dynamics such as exchange interactions that are extremely rare even in the densest areas of the field. It is thus reasonable to assume that the evolution of this pulsar should differ substantially from the others in Table~\ref{tab:dns}.  PSR~J1518+4904 is also a curious system, in that it has a moderate eccentricity, comparable to that of PSR~B1534+12, however, its measured transverse velocity is relatively low.  Most interesting is that the component masses seem to be relatively extreme, with a very light first-formed NS and a relatively heavy companion NS.  This, along with the system's relatively long orbital period, indicates that it may have proceeded through an entirely unique evolutionary scenario, which is not yet well understood \citep{jsk+08, wwk10}.
PSR~J1829+2456 may be a close cousin to PSR~J1518+4904, given its currently known mass limits, as well as its spin and eccentricity, which are similar to the latter system \citep{clm+04, clm+05}.  PSR~J1811$-$1736 has the most eccentric orbit in Table~\ref{tab:dns}, indicating it may have narrowly escaped disruption after the second SN event.  It is unlikely that reliable proper motion estimates will be obtained in the near future, given the large estimated distance to this pulsar, along with its relatively poor timing precision.  Such a measurement would help to better discern its formation history \citep{lcm+00, cks+07}.  Finally, PSR~J1906+0746 was originally thought to be a DNS system in which we see the low-mass young companion NS to an unseen recycled pulsar \citep{lsf+06, kas08}. The pulsar in this system is believed to be a young NS.  Following updated mass measurements from longer-term timing analysis \citep{kas12}, the nature of the companion is now ambiguous, and it may well be a massive WD.  In this case, it may have undergone an evolution similar to that of PSR~J1141$-$5645, another close NS-WD binary in a mildly eccentric orbit in which the massive WD was formed before the NS \citep{klm+00, ts00a, bbv08}. It has not yet been ruled out, however, that this system is a DNS.  If this is the case, an ECS may be the prevalent scenario for formation of the first-born NS; an ECS is also a potential channel for the second SN given the relatively low system eccentricity, although the companion mass may exclude this possibility \citep{kas12}.

One can see similarities between certain systems. In particular, PSR B1913+16 and PSR B1534+12 have massive companions, large eccentricities, and high transverse velocities. It is believed that these systems have undergone either a standard or double-He core-collapse SN of an iron core in forming the second NS.  Such high eccentricities and space velocities are indicative of a high mass-loss, asymmetric SN from a massive progenitor, giving a significant natal kick to the system  \citep{wkk00,wkh04,tds05,wwk10}. 

In contrast, the \psr{0737}A/B system has an order-of-magnitude smaller eccentricity and transverse velocity than those measured for the \psr{1534} and \psr{1913} systems \citep{sttw02,wnt10}.  This indicates there may have been little mass lost, and a small kick to the system from the second SN.
In addition, the mass of pulsar B is also significantly lower than any of the components in these two other DNS binaries.  Taken together, these properties suggest a different evolutionary path has been taken in the formation of pulsar B.

It has been suggested that this may best be explained by pulsar B having formed after an ECS \citep{pdl+05}.  
In addition to the low transverse velocity of the \psr{0737}A/B system, the mass of pulsar B provides important evidence for this. 
The mass corresponding to the binding energy of the NS for many equations of state is given by 
$E_{\mathrm{B}} \simeq 0.084(M_{\mathrm{NS}}/M_\odot)^2\msun c^2$ \citep{ly89,lp01}, corresponding to a mass $M_{\mathrm{BE}}\sim 0.13\msun$ for pulsar B \citep[see also][]{std+06}.
Through independent modeling of the pre-SN mass of the B pulsar based on a collapsing O-Ne-Mg core, \citet{pdl+05} have found that the critical mass for collapse should range from 1.366 to 1.375$\msun$.  Comparing with the measured mass of $1.249\msun$ for pulsar B \citep{ksm+06}, this is consistent with the above estimate for $M_{\mathrm{BE}}$.  This would mean that \emph{almost no baryonic matter has been lost} during the SN event.
Any remaining energy would have gone into changing the orbital properties of the system, or contributing to a kick at the time of the SN.  This is not only evidenced by the low measured projected velocity of the system, but also its small eccentricity.

Finally, we leave discussion of the \psr{1756} system, and how it compares to the double pulsar system, until Section \ref{sec:discussion}.  However, one can see that the component masses, as well as the orbital eccentricities, are very similar to those in \psr{0737}A/B.  It can thus be argued that this system may have followed a similar evolutionary path to that of the double pulsar system.  This would have important consequences for population studies of DNS systems.

\section{Unique Evidence for a Symmetric Supernova}
Together, the properties of the \psr{0737}B system present tantalizing clues that the progenitor of the B pulsar may have ended its life in an ECS, or other symmetric SN event.  However, it may still be possible to devise an evolutionary model that can reproduce its measured properties, especially given the possibility that a small transverse velocity does not necessarily imply a small velocity in the radial direction, and thus a small overall space velocity \citep{kvw08}.  As we now discuss, measurement of the spin direction of pulsar A can help to decide on the validity of an ECS or iron core-collapse scenario as the pulsar B formation mechanism.

It is expected from accretion theory that the spin axis of pulsar A will become aligned with the total angular momentum of the binary system (well approximated by the orbital angular momentum) as it accreted matter from the progenitor of pulsar B.  If the second SN is close to symmetric, this alignment would not be disturbed \citep{plp+04}. By contrast, if there is a large kick to the system, the resulting misalignment would equal the angular difference in orbital plane orientation before and after the SN event \citep[e.g.,][]{wkk00}.  
Several studies \citep[e.g.,][]{dv04,ps05,wkf+06,std+06, wwk10} have examined the explosion of the pulsar B progenitor.  Using the timing-derived proper motion from \citet{ksm+06}, \citet{std+06} predict a post-SN misalignment angle $\delta$ for \psr{0737}A of $\lesssim 11\degrees$.  \citet{wkh05} predict compatible values under different kinematic and progenitor-mass assumptions.

If the above studies are correct (see \citealt{kvw08} for a discussion and criticism of the assumption of a small radial velocity used in \citealt{ps05} and \citealt{std+06}), the measurement of a low spin-orbit misalignment angle for \psr{0737}A---in conjunction with the low pulsar B mass, system eccentricity, and transverse velocity that have been determined from timing measurements---would thus provide very favorable evidence 
for an ECS/symmetric SN event having formed pulsar B.
We thus aim to determine the orbital geometry of the \psr{0737}A/B binary in order to constrain the misalignment between its orbital angular momentum and the pulsar A rotation axes.  

One way to accomplish this is through investigation of the observable effects of geodetic precession on pulsar A.  In general, precession of the spin axis orientation of a pulsar causes a change in our line of sight through the emission region over time.  This is reflected in a continuous change in the observed pulse profile shape.  The extent to which these effects can be observed depends on the spin and orbital geometries of the pulsar system.  This is particularly the case for the misalignment angle $\delta$ between the pulsar spin and orbital angular momentum, which forms the opening angle of the cone that is swept out by the spin axis over the course of a precession period.  These geometric parameters can be modeled and determined through long-term analysis of the pulse profile.

The precession periods for the other relativistic DNS binaries mentioned earlier, \psr{1534} and \psr{1913}, are approximately 706 and 298 years, respectively.  Misalignment angles have been measured for these pulsars: $\delta = 25.0 \pm 3.8\degrees$ and $18 \pm 6\degrees$ for \psr{1534} and \psr{1913}, respectively \citep{wrt89,kra98,sta04}.
By comparison, \psr{0737}A has a precession period of 75 years 
\citep{lbk+04}.  The effects of geodetic precession on the pulse profile in this system should thus be more readily observable on a shorter timescale than for \psr{1913} and \psr{1534}, provided that $\delta$ is non-negligible.

In this article, we present a study of profile evolution for pulsar A over 6 years of observation.  After giving a brief overview of previous work in determining the geometry of pulsar A in Section \ref{sec:previous_work}, we describe our observations in Section \ref{sec:obs}. We outline our analysis and calculations of pulsar A profiles in order to obtain geometrical constraints in Section \ref{sec:profiles}.  In Section \ref{sec:results} we describe our results.  In Section \ref{sec:discussion} we discuss the implications of these results and compare them to previous findings. Finally, in Section \ref{sec:conclusions} we provide concluding remarks.

\begin{figure*}
  \begin{center}
    \includegraphics[width=0.49\textwidth]{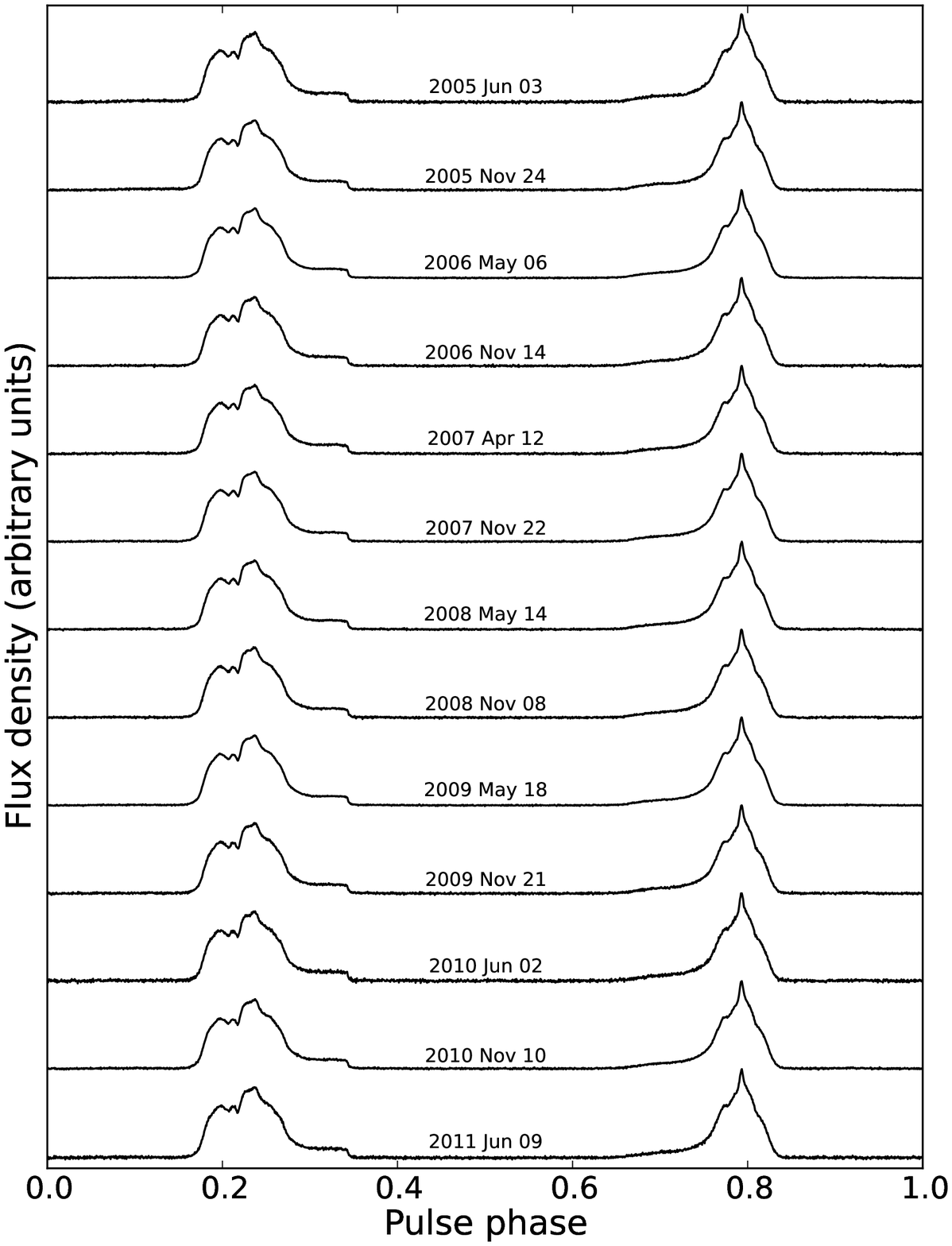}
    \includegraphics[width=0.49\textwidth]{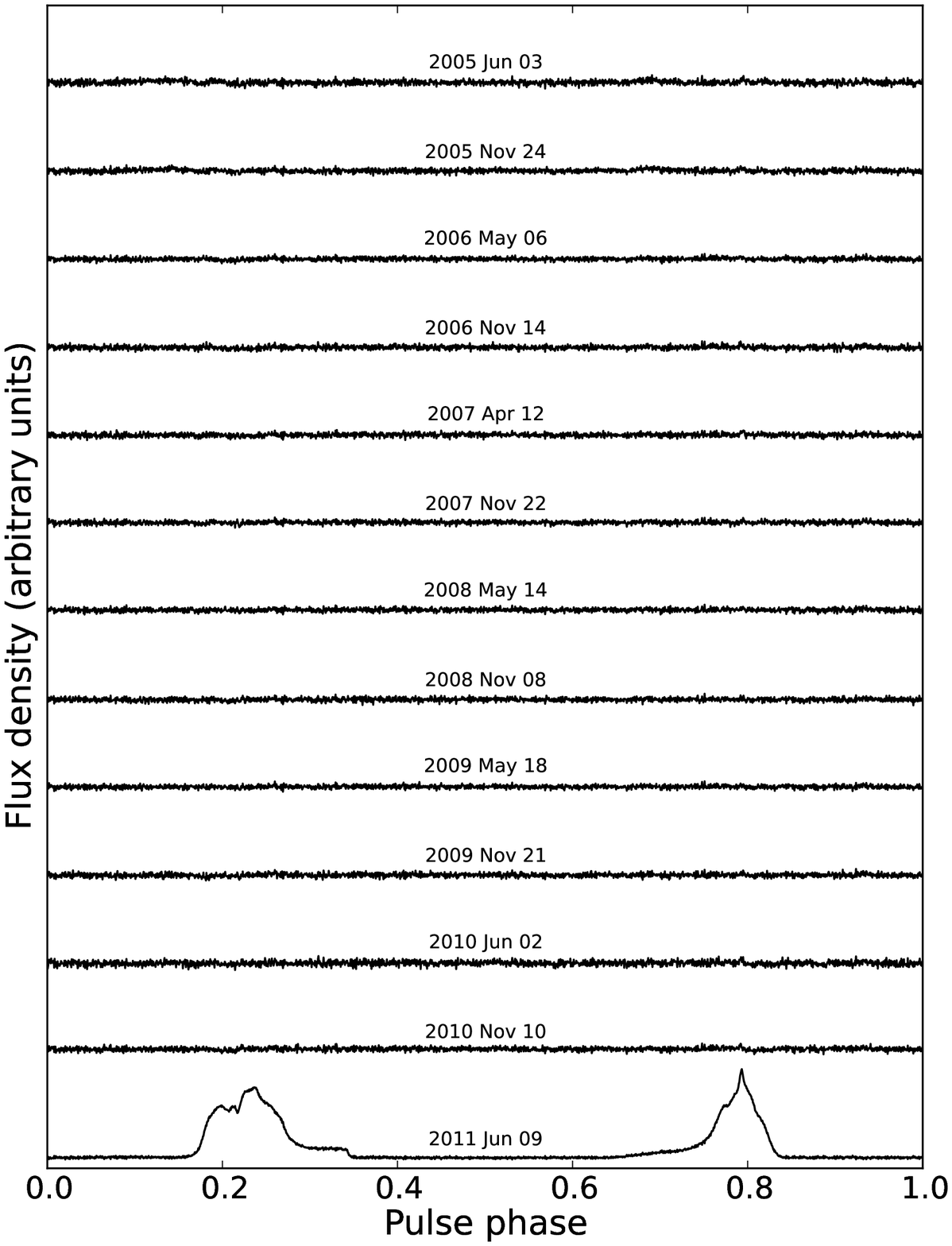}
    \caption{Integrated pulse profiles and difference profile residuals  over nearly 7 years of GBT observations.  Left: high signal-to-noise, aligned and averaged profiles are shown from each of 13 concentrated observing campaigns, and labeled with the central date of the given epoch.  Right: the most recent of these, taken in 2011 June, is subtracted from each of the other averaged profiles.  
The difference profiles in the right-hand panel are plotted at twice the vertical scale of the left-hand panel.\label{fig:profs}}
  \end{center}
\end{figure*}

\section{Previous Work}
\label{sec:previous_work}

Radio polarization data analysis was performed by \citet{drb+04} in order to derive geometrical constraints from position angle variations across the \psr{0737}A profile.  That work found solutions that prefer a spin-magnetic axis separation angle value of $\alpha \sim 5\degrees$ or $\alpha \sim 90\degrees$. They favor the former low-$\alpha$ solution, in order to avoid an emission beam opening angle $\sim 180\degrees$.  
However, \citet{drb+04} describe the changing position angle across the profile using the rotating vector model \citep[RVM;][]{rc69a}, which does not appear to be a good fit to the data 
(by the authors' own admission, the RVM provides an unsatisfactory description of the profiles of many recycled pulsars; see, e.g., \citealt{nms+97, xkj+98, stc99, yms+11}).

Using a data set that spans almost three years, \citet{mkp+05} found no evidence that the profile width of \psr{0737}A was changing with time. 
Because of the symmetric properties of the observed pulse profile, they interpreted both pulse components as emission from a single magnetic pole.  To avoid implying a beam opening angle of greater than $90\degrees$, they suggested a small but non-zero spin-orbit misalignment angle, with $\delta \sim 14\degrees$ as the preferred value, combined with a precessional phase close to either $0\degrees$ or $180\degrees$. A non-zero misalignment angle for pulsar A implies that pulsar B received a significant natal kick.
However, the subsequent precision timing analysis by \citet{ksm+06} and very long baseline interferometry measurement of its annual geometric parallax by \citet{dbt09} have both confirmed the proper motion, and thus transverse velocity, of the double pulsar system to be relatively low compared to other DNS systems.  This and the studies conducted by \citet{ps05}, \citet{wkf+06}, and \citet{std+06} suggest that a large natal kick is unlikely for the \psr{0737}A/B system, assuming the system has a small radial velocity. This would in turn predict a low value for $\delta$. 

If \psr{0737}A is a near-orthogonal rotator, then the profile we observe from this pulsar may represent contributions from each of the two magnetic poles.  
Additionally, the two pulse components are separated by approximately $180\degrees$ and the region between is largely free of pulsed emission.  With this model there is not a problem in postulating a small or even zero value of the spin-orbit misalignment angle.  A detailed study of the pulse shape and polarization, and investigation of possible emission mechanisms is necessary to achieve a more satisfactory understanding of the geometry of the system. 

To further investigate, we have conducted a similar analysis to \citet{mkp+05}, performing a fit of the geometrical parameters describing the \psr{0737}A/B system to the pulse width data.  We have extended our data set to include 6 years of observations of this system, making the constraints on the system geometry far more rigid 
\footnote{A preliminary version of this analysis can be found in \citet{fer08} and \citet{fsk+08}}. 
As we describe below, we perform this investigation assuming both one- and two-pole beam models; as we discuss in Section \ref{sec:discussion}, we find that the latter emission structure allows for a smaller beam size, and is our preferred description for this system.

\section{Observations}
\label{sec:obs}

Observations of \psr{0737}A were taken between MJD 53521 (2005 May 31) to MJD 55731 (2011 June 19) using the 100-m Robert C.~Byrd Green Bank Telescope (GBT) in West Virginia, with the Green Bank Astronomical Signal Processor backend \citep[GASP;][]{dem07}.  GASP is a flexible baseband system, which performs 8-bit Nyquist sampling of the incoming data stream at $0.25 \mu$s intervals in both orthogonal polarizations.  The incoming data stream was then coherently dedispersed \citep{hr75} in software.  After this, the signals were folded at the topocentric pulse period of PSR~J0737$-$3039A to form pulse profiles, each of which were accumulated over approximately $30\,$s. This choice of integration time reflects a compromise between ensuring a fully-sampled orbit for the very precise Shapiro delay measurement for timing studies, and obtaining pulse profiles with adequate signal-to-noise ratio to effectively carry out that analysis.

These folded pulse profiles were typically flux-calibrated in each polarization using a reference signal from a noise diode source that was injected at the receiver.  When calibration data were not available, we normalized the profile data in each polarization by the root mean square (rms) value of the corresponding off-pulse signal.  The data were finally summed over both polarizations and across all frequency channels to give the total power signal for each integration \citep[for further details on GASP operation and data reduction, see][]{dem07,fer08}.

As stated above, the observing strategy for the double pulsar has been such that it would benefit the timing analysis that has produced the most recent stringent test of GR \citep{ksm+06}.  To that end, we have been taking data in monthly observing sessions, each consisting of a 5--8 hr track of the object.  The incoming signal was divided into a maximum of $16$ and $24 \times 4\,$MHz channels\footnote{The number of channels used occasionally varied due to radio frequency interference (RFI) and available computing resources.}, centered at 820 and 1400 MHz, respectively, in alternate months.  This was done in order to effectively constrain astrometric parameters such as position and proper motion, as well as possible changes in the measured dispersion measure (DM) of the pulsar.  
We have also conducted semi-annual concentrated observing campaigns that cover approximately 1--2 weeks each, and which typically contain up to six observing sessions that each last 5--8 hr.  The purpose of these was to better constrain orbital parameters and to investigate the effects of special relativistic aberration on the pulse shape over an orbital period.  These campaign observations were taken exclusively using the 820-MHz receiver.  

The data set used for this work consists solely of these concentrated campaign observations.
We have excluded the Parkes telescope data set used in \citet{mkp+05}, as they were taken with a different observing system, at a different frequency. However, given that there is also a lack of change in those pulse profiles, the upper limits derived for $\delta$ in this work can be considered to be conservative.

\section{Long-Term Profile Analysis}
\label{sec:profiles}
In order to perform a long-term profile shape analysis, we first combined the integrated, calibrated pulse profiles from each set of campaign observations described above into high signal-to-noise profiles.  Each of these averaged profiles represented the midpoint in time during which each corresponding set of observations were taken.  The left panel of Figure~\ref{fig:profs} shows the combined average profiles from each representative epoch.  
For illustration, the right panel of Figure~\ref{fig:profs} shows the residual profiles from each corresponding epoch, obtained when the profile representing the data taken around MJD 55721 (2011 June 9) is subtracted from each of the other mean profiles.  
A qualitative glance at these difference profiles shows the lack of substantial changes between epochs.  Furthermore, there seems to be no secular change over time, which one may otherwise expect for this highly relativistic system, and which is the case for both \psr{1913} and \psr{1534}.
This would support the hypothesis of pulsar A's angular momentum being aligned with that of the system.

\subsection{Width Calculations}
\label{sec:widths}

After fitting for the pulse height with a simple Gaussian to several points around the profile peak, we calculated the widths and width uncertainties of the pulsar A profile at each epoch, at several fractional pulse heights. 
We did this through a bootstrap-type method, as follows: we first performed a polynomial interpolation to find the pulse phases on the profile, at the chosen fractional pulse height.  This was done for each side of the profile (or profile component, in the two-pole case) by randomly choosing 11 out of 25 possible (and not necessarily contiguous) data points along the profile, about the location in phase it is expected to cross the given height.  The difference between the fit phase on each side of the profile was taken as the calculated pulse width.  After 32768 such iterations, we constructed a histogram of these trial widths, to which we fit a Gaussian profile.  The mean value and $1\sigma$ width of this distribution were taken as the final pulse width and uncertainty for the given epoch, respectively.

To include the possibility that we may be seeing two-pole emission from this pulsar,  
we also treated each component, separated from each other by a region of negligible pulsar emission, as the distinct emission from each magnetic pole, and separately measured the width of each of these as described above.  

\begin{figure}
  \begin{center}
    \includegraphics[width=0.5\textwidth]{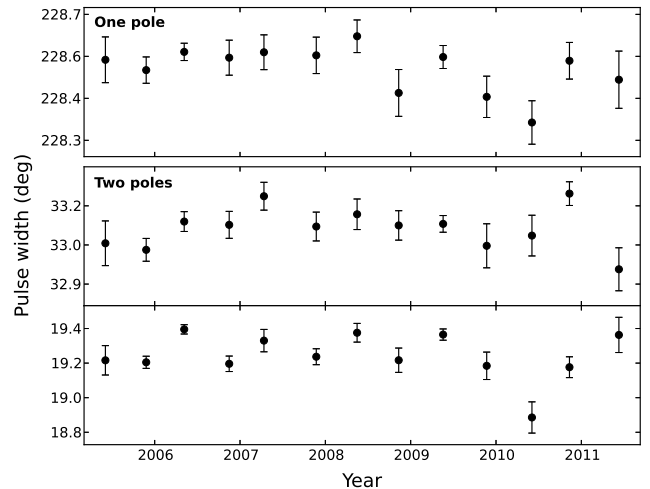}
    \caption{Width measurements for PSR~J0737$-$3039A at $40\%$ of the pulse height. The top panel shows the widths when assuming a one-pole emission model for this pulsar.  The bottom two panels show the measurements in the case of a two-pole model, where each profile component is assumed to originate separately from each beam.  For this case, the middle and bottom panels correspond to the first and second components, respectively.\label{fig:widths}}
  \end{center}
\end{figure}

To ensure that we obtained a reliable fit, we measured the profile widths at $5\%$ fractional height increments, between $30\%$ and $50\%$ of the total pulse height.  Within this region, there are no features (e.g.~the plateau region in the first pulse component at $\sim 10\%$ pulse height) that may affect our analysis from obtaining a clear determination of the pulse width.
As an example of our width measurements, Figure~\ref{fig:widths} plots those made at $40\%$ of the pulse height for each of the one- and two-pole emission cases. At a glance,  we see no perceivable \emph{secular} change in the pulse shape.  Any perceived outlying measurements are likely due to several possible effects; principal among these are likely low-level radio frequency interference (RFI), and lack of accounting for the non-orthogonal polarization feeds in our calibration \citep[see, e.g.,][]{van06,van12}. We performed the geometrical model fit using the set of measurements at each of the fractional pulse heights for which we have calculated widths.

\subsection{Geometry Fit}
\label{sec:fit}

In order to model the geometry of PSR~J0737$-$3039A/B, we use the formulation given by \citet{rl06a}, which relates the long-term profile evolution of pulsar A to the system geometry.  The dependence of the observed pulse width on the system geometry can be expressed by the following:
\begin{equation}\label{eqn:phi0}
  \cos{\Phi_0} = \frac{\cos\rho - \cos\zeta\cos\alpha}{\sin\zeta\sin\alpha},
\end{equation}
where $\Phi_0$ is the half-pulse width, $\alpha$ is the angle between the spin and magnetic axes, $\zeta$ is the angle between the spin axis and the observer line of sight, and $\rho$ is the half-opening angle of the part of the emission cone corresponding to the given fractional pulse width.  In using this equation, we are assuming a circular emission beam.  
In their work, \citet{mkp+05} investigated the effect of using a noncircular beam on their analysis, and found that the results were only marginally affected.  We thus believe that a circular emission beam is a reasonable assumption for this analysis. 
For schematic diagrams that show the angles involved in pulsar spin and orbital geometries in the above expressions, refer to Figure~1 in \citet{dt92}.

\begin{figure}
  \begin{center}
    \includegraphics[width=0.5\textwidth]{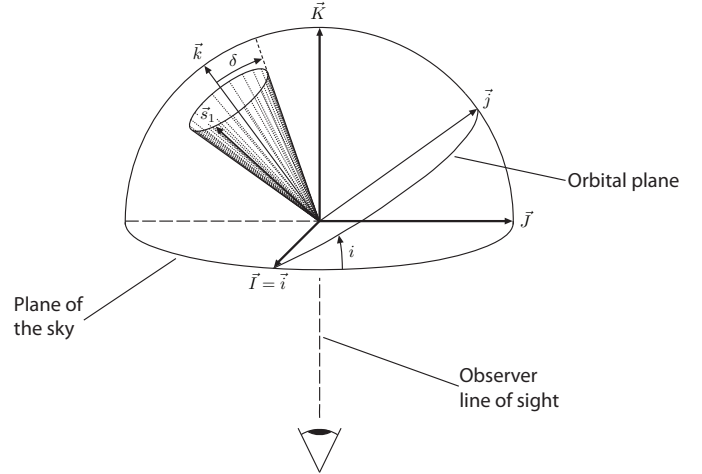}
    \caption[Diagram of pulsar precession geometry]{Pulsar precession geometry for a general system.  The vectors $\vec{I}$ and $\vec{J}$ define the plane of the sky, so the $\vec{K}$ points in the direction away from the observer. $\vec{i}$ and $\vec{j}$ define the plane of the pulsar orbit, so that $\vec{k}$ points in the direction of the orbital angular momentum. $i$ is the orbital inclination relative to the observer line of sight.  Due to geodetic precession, the pulsar spin axis $\vec{s}_1$ will trace out a cone with opening angle $\delta$, the misalignment angle between the pulsar spin axis and the orbital angular momentum. This diagram is adapted from \citet{dt92}.\label{fig:precession_geometry}}
  \end{center}
\end{figure}

In relativistic systems, we would expect that geodetic precession would have the general effect of varying the angle $\zeta$ over time.  In order to express Equation~\ref{eqn:phi0} in terms of parameters that do not vary with time, we make use of the fact that the direction of the spin axis vector $\vec{s}_1$ can not only be described by the polar angles $(\zeta,\eta)$ (where $\eta$ is the angle between ascending node and the projection of the spin axis on the plane of the sky), but alternatively by $(\delta,\phi_{\mathrm{SO}})$, assuming that the precession phase varies linearly in time.
Here, $\delta$ is the spin-orbit misalignment angle, and $\phi_{\mathrm{SO}}(t)$ is the longitude, or phase, of $\vec{s}_1$ in its precession around the orbital angular momentum vector $\vec{k}$, as measured from the $-\vec{J}$ axis, as illustrated in Figure~\ref{fig:precession_geometry} \citep{dt92}.  

The transformation $(\zeta,\eta)\rightarrow(\delta,\phi_{\mathrm{SO}})$ is given by \citet{dt92}:
\begin{align}
  \cos{\lambda} & = \cos{\delta}\cos{i} - \sin{\delta}\sin{i}\cos{\phi_{\mathrm{SO}}}\\
  \cos{\eta} & = \frac{\sin{\delta}\sin{\phi_{\mathrm{SO}}}}{\sin{\lambda}},
\end{align}
where $\lambda = \pi - \zeta$, and so $\sin{\lambda} = \sin{\zeta}$ and $\cos{\lambda} = -\cos{\zeta}$.  We thus have:
\begin{align}
  \label{eqn:zeta_to_delta}
  \cos{\zeta} & = -\cos{\delta}\cos{i} + \sin{\delta}\sin{i}\cos{\phi_{\mathrm{SO}}}\\
  \label{eqn:eta_to_delta}
  \cos{\eta} & = \frac{\sin{\delta}\sin{\phi_{\mathrm{SO}}}}{\sin{\zeta}}.
\end{align}
We can also recast $\phi_{\mathrm{SO}}$ as:
\begin{equation}
	\phi_{\mathrm{SO}} = \Omega_1^{\mathrm{spin}}(t - T_1),
\end{equation}
where $\Omega_1^{\mathrm{spin}}$ is the angular precession frequency, $t$ is the epoch, and $T_1$ is the reference crossing time of the spin axis through precession phase $\phi_{\mathrm{SO}}=0\degrees$. 
This analysis was performed separately for the cases of $\cos{i}>0\ (i=88.7\degrees)$ and $\cos{i}<0\ (i=91.3\degrees)$ \citep{ksm+06}.  We use a single value for the inclination in our analysis for each of these two cases; the fractional uncertainty in these values is small enough that doing so would not significantly affect our results.

Finally, ($\zeta$, $\eta$) are now given by:
\begin{align}
\label{eqn:zeta_delta_t1}
  \cos{\zeta} & = -\cos{\delta}\cos{i} + \sin{\delta}\sin{i}\cos{[\Omega_1^{\mathrm{spin}}(t - T_1)]}\\
\label{eqn:eta_delta_t1}
  \cos{\eta} & = \frac{\sin{\delta}\sin{[\Omega_1^{\mathrm{spin}}(t - T_1)]}}{\sin{\zeta}}.
\end{align}

Substituting Equation~\ref{eqn:zeta_delta_t1} into Equation~\ref{eqn:phi0} gives us an expression for the time-varying pulse width $\Phi_0$, and that depends on the three angles $\alpha, \delta,\mbox{ and } \rho$, as well as the epoch of zero precession phase $T_1$.  

These parameters can be determined through a least-squares fitting of our pulse width data to the model given by Equation~\ref{eqn:phi0}.  Here, we have performed a three-dimensional grid-search fit, over the ranges $0\degrees < \alpha < 180\degrees$, $0 < \delta < 90\degrees$, and with $T_1$ running over one precession period centered on 1990 January 1 (MJD 47892).  Although it is physically and mathematically possible for $\delta$ to have values between $90\degrees$ and $180\degrees$, we focus on values $< 90\degrees$; it has been shown by \citet{bai88} that, based on modeling of \psr{1913}, misalignment angles with these values are much more physically likely, unless the SN kick was extremely large.

\begin{figure*}
  \begin{center}
    \includegraphics[width=1.0\textwidth]{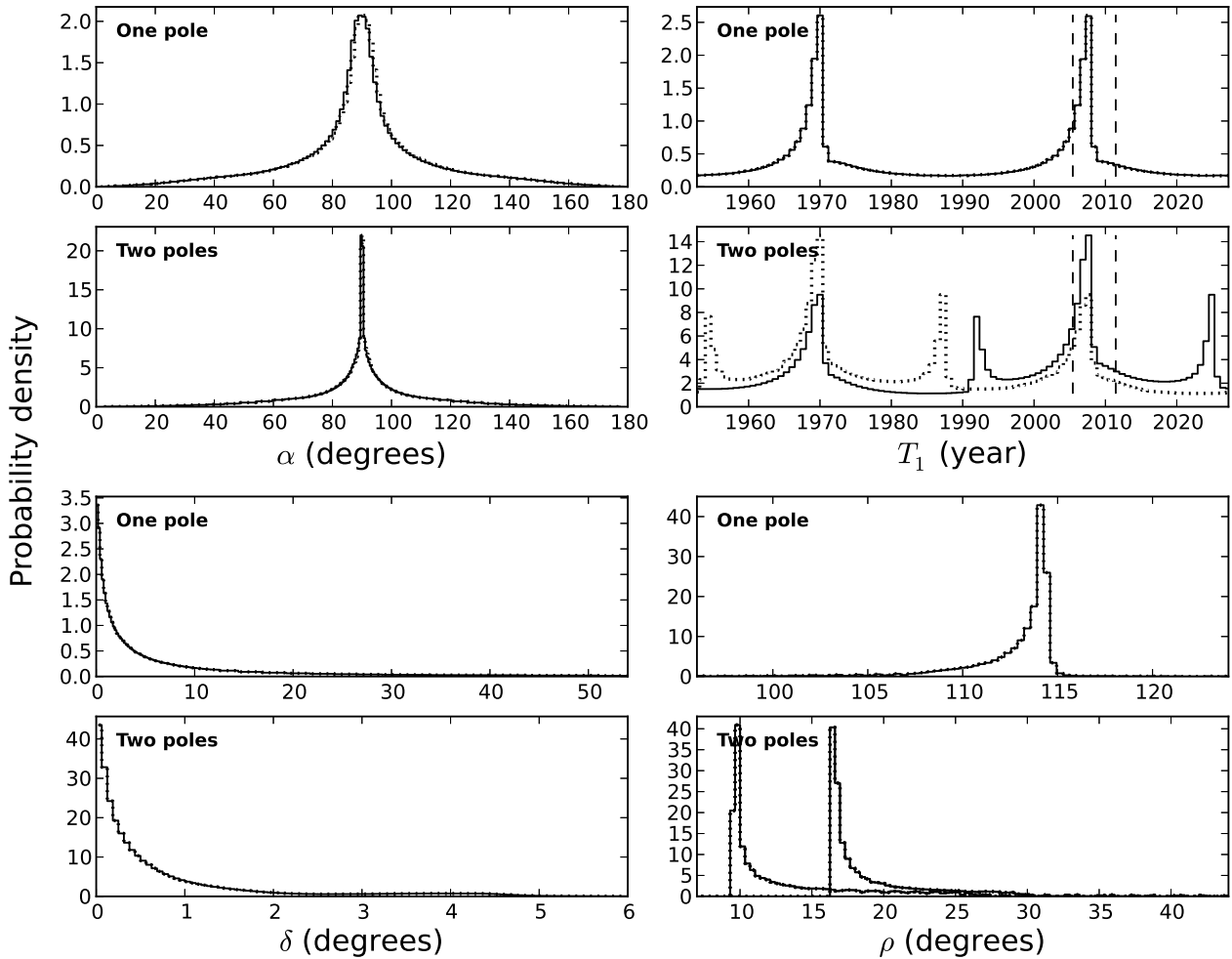}
    \caption{Probability density functions (PDFs) for each of the geometric parameters in the model of \citet{rl06a} for the PSR~J0737$-$3039A/B system, found using widths calculated at a $40\%$ fractional pulse height.  Results for the case of orbital inclination $i<90\degrees$ are represented by solid lines, and those for $i>90\degrees$ by dotted lines.  Shown are the PDFs for the angle $\alpha$ between the spin and magnetic axes of pulsar A (top left), the misalignment angle $\delta$ between the pulsar A spin axis and the total angular momentum of the system (bottom left), the epoch $T_1$ of zero geodetic precession phase (top right), and the half-opening angle $\rho$ of the region of the pulsar A beam that corresponds to the measured width at $40\%$ of the total pulse height (bottom right). The results for the one- and two-pole cases are shown in the top and bottom sub-plots for each parameter, respectively. The vertical dashed lines in the $T_1$ PDF plots represent the range of our data span.\label{fig:pdfs}}
  \end{center}
\end{figure*}

For each fractional pulse height, we run a Levenberg-Marquardt fit of the corresponding width data to the above model at each grid point, allowing only $\rho$ to vary.  In the one-pole model case, there is only one beam and thus a single $\rho$ value found at each grid point.  In the two-pole case, for each fractional pulse height, we have two width data sets that we fit simultaneously into the single model described above.  We make the assumption that the magnetic poles are separated by $180\degrees$, so that we substituted $\alpha \rightarrow 180\degrees - \alpha$ in the model when describing the second component in the pulsar A profile.  Here, there are two emission beams and thus two distinct $\rho$ values for which to fit (henceforth referred to as $\rho_1$ and $\rho_2$, corresponding to the first and second profile components, respectively).

\input{0737A_geometry_results_table_arXiv}

After performing a fit over a coarsely sampled grid in each dimension, we did a final fit with variable grid-point bin sizes, such that regions of higher probability density were re-sampled with finer bins, and regions with relatively little or negligible probability density were probed with coarser sampling.
This was done to provide an accurate calculation of probability density, while minimizing computation time.  
We then arrived at a joint probability density distribution for $\alpha$, $\delta$, and $T_1$, from which we found individual probability density functions (PDFs) for each of these parameters in turn by marginalizing over the other two parameters.  These fits and PDF calculations were done for both the one- and two-pole emission models, for each set of fractional pulse height width data, and for the two possible inclination values: $i < 90\degrees$ and $i > 90\degrees$.  In Figure~\ref{fig:pdfs}, we show the resulting PDFs for the one- and two-pole model fits to the width data taken at $40\%$ of the peak pulse height. We found PDFs for $\rho$ by calculating a histogram of all fit $\rho$ values, weighted by the output probability density at the corresponding grid point.  Results for both orbital inclinations are very similar, which is unsurprising, given how close the true inclination is to $90\degrees$.  Even with independent measurement of any of the geometrical parameters, it would in practice be very difficult to discern which inclination is correct; such a measurement would have to be extremely precise.  In what follows, we will refer solely to the results obtained for the case of $i < 90\degrees$.

\section{Results}
\label{sec:results}

We present the fit geometric parameters in Table~\ref{tab:geometry_result}, where our results are shown for each of the one- and two-pole models, and for each fractional pulse height at which we calculated the input width data. As stated above, we only show the results for $i < 90\degrees$.
The median and $68\%$ confidence intervals for the histogram are used as the measured $\alpha$ and $\rho$ values and their corresponding uncertainties, respectively.  For $\delta$, we have determined $68\%$, $95\%$, and $99\%$ upper limits.  

The one- and two-pole emission cases give consistent results for $\alpha$ and $\delta$, favoring the pulsar A axis of rotation to be at approximately right angles to the magnetic axis, and a low spin-orbit misalignment angle. However, a major difference between the two emission models comes with the preferred values we find for $\rho$. Since $\rho$ ($\rho_1$ and $\rho_2$ in the two-pole case) is the beam half-opening angle corresponding to the given fractional pulse width, it is always smaller than the half-opening angle of the full beam. The values we find for $\rho$ in the one-pole case thus greatly favor a full beam opening angle that is larger than $180\degrees$.  The two-pole case, however, easily allows for two emission beams on opposite sides of the NS, each with an opening angle significantly less than $180\degrees$. 
Allowing emission to come from both magnetic poles in this pulsar thus avoids the need for exotically large beams, or a reinterpretation of the the location of the emission beam center.

It is for this reason that we strongly favor geometry such that this pulsar is a \emph{near-orthogonal rotator emitting from both magnetic poles, and with the spin axis closely aligned with the orbital angular momentum vector}.  From this point, we will thus use the results from the two-pole case in our discussion.  We quote an average value for the spin-magnetic axis separation of $\alpha = 90\pm11\pm5\degrees$, where the first uncertainty quantifies statistical effects, calculated as the median value of the determined $\alpha$ uncertainties.  The latter represents that which is due to systematic error (likely due to RFI and calibration effects as described in Section \ref{sec:widths}), and is taken to be the difference between the maximum absolute uncertainty and the statistical error.  We also quote $68\%$, $95\%$, and $99\%$ average upper limits to  the pulsar A spin-orbit misalignment angle $\delta$ of $0.85\degrees, 3.2\degrees$, and $4.7\degrees$, respectively. 

The epoch of zero precession phase $T_1$ is difficult to constrain, given the geometry of the system. It seems to favor values such that $\phi_\mathrm{SO} = 0\degrees$ or $180\degrees$ falls within the span of our data set, denoted in Figure~\ref{fig:pdfs} by vertical dashed lines.
However, for $\delta \sim 0\degrees$, any fit value for $T_1$ would be fairly meaningless, since it would be impossible to define a precession phase at any given time.  Unless the geometrical parameters are very finely tuned, if pulsar A had a measurable misalignment angle $\delta$, we would expect to see at least \emph{some} noticeable evolution in the pulse profile shape in the 7 years ($>8\%$ of its precession period) we have been monitoring this pulsar with the GBT.  This is an especially valid argument, given the significant changes one sees in other DNS profiles whose precession periods are far longer than that of \psr{0737}A.  It is for these reasons that we do not include $T_1$ in Table~\ref{tab:geometry_result}.

\begin{figure}
  \begin{center}
%%% set at 0.8(textwidth) here to avoid losing some of caption off 
%%% the bottom edge in manuscript layout
    \includegraphics[width=0.5\textwidth]{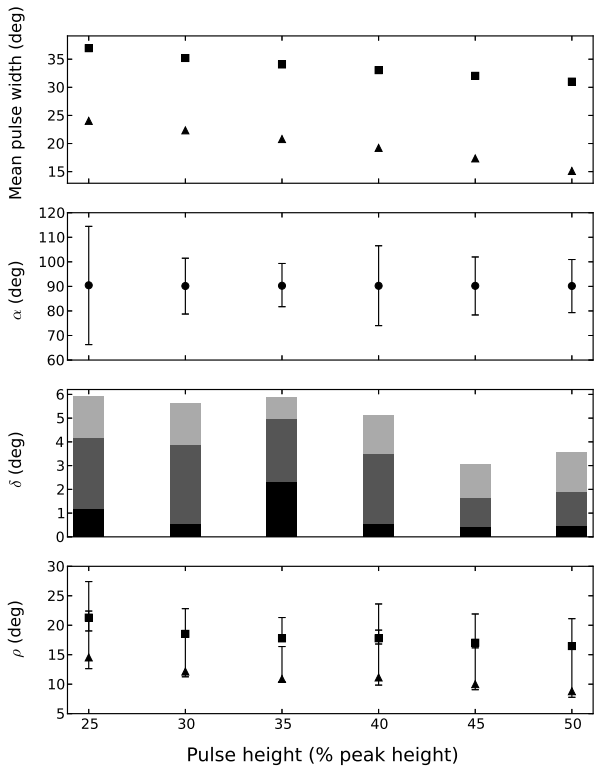}
    \caption{Summary of results of J0737$-$3039A profile analysis assuming two-pole emission, for $i<90\degrees$.  Each set of results is plotted against the corresponding fractional pulse height for which they were found.  Top panel: mean pulse width.  Uncertainties here are smaller than the plotting symbol sizes.  Second panel from top:  median value for $\alpha$, with $1\sigma$ uncertainties.  Third panel from top:  misalignment angle $\delta$. Here, the darkest to lightest regions represent the $68\%$, $95\%$, and $99\%$ upper limits, respectively.  Bottom panel:  median values of $\rho$, with $68\%$ uncertainties.\label{fig:results_summary}}
  \end{center}
\end{figure}

Figure~\ref{fig:results_summary} summarizes these results for the $i < 90\degrees$ case.  One can see that the results for all the fits we perform are consistent across all width data sets.
Finally, regardless of which of the one- or two-pole emission models one prefers, we can see that the favored geometry for this system is one with a pulsar A spin axis at near right angle to its magnetic axis, with a very small spin-orbit misalignment.

\section{Special Relativistic Aberration}
\citet{rl06a} also suggested that differential delays due to latitudinal aberration should modify the pulse profile of pulsar A on an orbital timescale \citep[see also][]{dt92, sta04}. This aberration arises from the fact that the line of sight for observers in different inertial frames intercepts the emission beam at different co-latitude angles. Hence, the emission components located on either side of a hollow beam are expected to move towards and away from each other as the sight line intercepts the beam closer to the edge or the center, respectively. The orbital motion should therefore induce a periodic change of the latitudinal aberration, which will cause a sinusoidal variation in the separation between the pulse components. Thus, a given component of the emission beam will suffer a shift having an amplitude proportional to the orbital velocity multiplied by a geometric factor which depends on the beam and spin axis geometry: 
\begin{eqnarray}
\label{eqn:aberration}
\Delta \Phi_0  = \frac{\beta}{\sin \zeta \tan \chi_0} & [ \cos i \sin \eta (\cos\psi + e\cos\omega) \nonumber  \\
                                                     & - \cos\eta (\sin\psi + e \sin\omega) ], 
\end{eqnarray}
where $\beta$ is the orbital velocity, $\zeta$ and $\eta$ have been defined in Section \ref{sec:fit}, $\psi$ is the orbital phase, $e$ is the orbital eccentricity, $\omega$ is the longitude of periastron, and
\begin{equation}
\tan\chi_0 = \frac{\sin \alpha \sin \Phi_0}{\cos \alpha \sin \zeta - \cos \Phi_0 \sin \alpha \cos \zeta}, 
\end{equation}
where in this case $\Phi_0$, defined as in Equation~\ref{eqn:phi0}, is the spin longitude as measured from the beam center, and the other quantities here are defined earlier in this paper.

If we believe that the two symmetric components of pulsar A's profile are the leading and trailing edges of a single pole, then they will suffer a periodic shift $2 \Delta \Phi_0$. In the case of a two-pole model, one can show that under a symmetry assumption of the emission beams the shift will also be the same, though the definition of the beam geometry (e.g. $\alpha$, $\rho$, etc.) will apply to the two poles. From this, it follows that one can constrain pulsar A's geometry by measuring the separation between the two pulse components as a function of orbital phase. 
We have performed this analysis using observations at the GBT, both with data from the SPIGOT \citep{kel+05} and GASP backends.
As is the case for the long-term evolution of the pulse profile, we found no significant change in the separation of the pulse components. We can set an upper limit to the value of the denominator of Equation~\ref{eqn:aberration}, $(\sin \zeta \tan \chi_0)^{-1}$, of 0.13, 1.33 and 4.68, corresponding to the $1\sigma$, $2\sigma$ and $3\sigma$ confidence levels, respectively. Given that the equations for this analysis are similar to those found in Section \ref{sec:fit} for the long-term profile evolution, the implications for the geometry of pulsar A are similar to those presented earlier, though less stringent.

\section{Discussion}
\label{sec:discussion}
These results are somewhat at odds with previous work.  
Through position angle variation analysis, \citet{drb+04} prefer solutions with $\alpha \sim 5\degrees$ or $\alpha \sim 90\degrees$. They choose the former low-$\alpha$ solution, in order to avoid an emission beam opening angle of approximately $180\degrees$.
Although the analysis by \citet{mkp+05} favors $\delta \lesssim 60\degrees$, consistent with our findings, they believe that a spin axis aligned with the orbital angular momentum is unlikely.  This was due to early scintillation-based measurements of the transverse velocity of \psr{0737}A/B, which supported a large natal kick \citep{rkr+04,wkh04,cmr+05}. As discussed earlier, however, more recent and reliable timing measurements confirm a transverse velocity for the system of less than $\sim 10\kms$. Furthermore, \citet{mkp+05} arrive at best-fit values for $\alpha$ that virtually exclude $\alpha\sim 90\degrees$.  This is in direct contrast to our findings, which strongly disfavor low-$\alpha$ solutions (as well as those near $180\degrees$). 
As with \citet{drb+04}, \citet{mkp+05} enforce a half-beam opening angle of less than $90\degrees$ in their model of the pulse width data, reasoning that values larger than $90\degrees$ are unphysical, or else means that the beam center has been misidentified.  
We have shown, however, that such a large beam can easily be avoided by allowing the emission to originate from both magnetic poles.

Although we strongly prefer a two-pole model for polar cap emission from pulsar A, it has also been postulated that the observed wide pulse profile in many recycled pulsars is due to emission that originates in the outer magnetosphere.  Here, caustic effects produce a fan-shaped beam that, depending on its orientation, could be viewed with an opening angle that is greater than $180\degrees$ \citep{rmh10}.  The observed coincidence of millisecond pulsar pulse profiles with counterparts found in observations with the \textit{Fermi} Large Area Telescope (LAT) seems to support this idea for certain cases \citep[e.g.,][]{aaa+10b, egc+12arxiv}. Such a model may also explain the frequent disagreement of recycled pulsar polarization data with a standard RVM description \citep{yms+11}; for pulsar A in particular, it provides reasoning for the observed symmetry in its pulse profile.  
However, analysis of \textit{Fermi} LAT observations of \psr{0737}A/B clearly shows that the $\gamma$-ray and 1.4-GHz radio data are \emph{not} aligned in rotational phase.  This, along with geometric modeling that uses both $\gamma$-ray and radio data, support a two-pole emission model (L.~Guillemot et al. 2013, in preparation).
Regardless of the choice of emission model, our main result of nearly aligned spin and orbital angular momentum axes for pulsar A remains unchanged. However, the outer magnetosphere emission model described above would give more relaxed constraints on the system geometry, similar to the one-pole emission model results shown in Table~\ref{tab:geometry_result}.

A low spin-orbit misalignment angle means that a very small kick was likely given to the \psr{0737}A/B system due to the SN that resulted in pulsar B.  This 
is supported by measurements of a low transverse velocity and small eccentricity for this system.  
The low mass of pulsar B also lends credence to the studies performed by \citet{ps05}, \citet{wkf+06}, and \citet{std+06}, which favor a low pre-SN mass ($< 2\msun$) progenitor, and a low natal kick ($\lesssim 100\kms$) given the constraints of a low radial velocity.  In the case of the latter study, our constraints are also consistent with their estimate of the misalignment angle, which they predict to be $0.5 \le \delta \le 11\degrees$ ($95.4\%$ confidence).  

This analysis thus supports a scenario in which the \psr{0737}B progenitor underwent a low mass-loss, relatively symmetric SN event.
The prominent candidates for such an event are an ECS, or possibly the collapse of a low-mass iron core \citep{pdl+05,plp+04}, and we therefore favor one of these scenarios as the one that produced the double pulsar system as we now observe it.  Discovery and observations of an increasing number of DNS binary systems will help to determine the prevalence of this type of DNS evolutionary channel.

In addition, these new results provide an impetus for updating various studies that aim to constrain the kick velocity given to the \psr{0737}A/B system after the second SN.  For example, the simulations done by \citet{std+06} can now be improved upon, by including our upper limits on $\delta$ as a prior distribution in their analysis.  In the work done by \citet{wwk10}, tighter constraints can also be achieved by using our $\delta$ values, particularly if one adopts the two-pole emission model, as we believe is correct.  The method of the latter study also holds the potential for constraining the radial velocity of this system.  This is crucial if we are to once and for all determine whether the \psr{0737}A/B system velocity has a large radial component, the deciding factor in distinguishing between space velocities, and thus the magnitude of the natal kick on the system after the second SN.  

These results may also help to determine the cause of the very large misalignment angle for pulsar B reported by \cite{bkk+08}, and more broadly, calls into question the origin of the pulsar B spin.  
For example, noting that the majority of the currently observed pulsar B spin must have come from the second SN, \citet{fklk11} has addressed this issue by postulating that any natal kick resulting from the SN that formed pulsar B must have been displaced from the center of mass of the pulsar B progenitor star.  The measurement of the spin-orbit misalignment in pulsar A presented here may contribute to tighter constraints on the magnitude of this displacement in such a model.

Finally, we speculate on the possibility of other DNS systems having been formed through an ECS (or symmetric SN event) of the second-formed NS progenitors in these systems.  If we look back at Table \ref{tab:dns}, we see distinct similarities between several measured properties of PSRs J1756$-$2251 and those of the double pulsar system.  This is a close DNS systems, with a low eccentricity and a low second-formed NS mass.  It has been argued \citep[e.g.,][]{vdh04,vdh07,wwk10} that the evolutions of this, and possibly other DNS systems, may also have undergone ECS events in their histories.

PSR~J1756$-$2251 is the recycled, first-formed pulsar in that system.  Long-term pulse profile monitoring can thus be done to constrain its geometry \citep{fer08}.  Due to its proximity to the ecliptic, the determination of its transverse velocity has thus far presented a challenge, however, the measurement of the right ascension component of its proper motion corresponds to a velocity of $9.3\kms$ in that direction, based on a DM distance of 2.5 kpc for this pulsar \citep{cl02, fkl+05, fer08}.  This low measured velocity presents another possible clue in favor of the \psr{1756} system having undergone a similar evolution to the double pulsar.  It is hoped that future observations will elucidate whether this is indeed the case.

It may be that ECS (or another small-kick SN scenario) may be the dominant channel for close DNS formation \citep{plp+04}---the likelihood of a system remaining bound after \emph{two} SNe is more likely if one of these involves a low mass-loss, small-kick event.  On the other hand, the narrower range of the progenitor masses required for this channel to proceed may reduce the population of suitable systems for this scenario.

\section{Conclusions}
\label{sec:conclusions}

In this article, we have presented the calculation of constraints on the geometry of the \psr{0737}A/B system through measurement of high signal-to-noise integrated profiles of pulsar A.  Our results favor an interpretation where pulsar A's magnetic axis is nearly perpendicular to the rotation axis and where the system emits from both magnetic poles as a near-orthogonal rotator.
Most interestingly, we find that the angle $\delta$ between the pulsar A spin axis and the total angular momentum of the system is small, consistent with an alignment between the two axes.  This provides very important evidence in favor of pulsar B having formed as a result of a symmetric event, possibly an ECS.  

Further discovery and observation of other DNS systems will help to determine how frequent such an evolution may be as a channel for DNS system formation.  Gaining a better understanding of the number of DNS systems that proceed through different formation schemes will be important for population synthesis analysis, and for predictions of possible gravitational wave (GW) sources for upcoming detectors such as Advanced LIGO \citep{ligo10} and Advanced Virgo \citep{virgo11}, which will be sensitive to GWs from DNS mergers.

A final qualitative conclusion to be drawn from this work is that the remarkable long-term stability of the \psr{0737}A pulse profile indicates that current and future timing analysis performed through observations of this pulsar is likely to be extremely reliable over a long time span.  This is especially important for new and continued tests of GR that will be undertaken with this astonishing system.

\acknowledgements
Pulsar research at UBC is supported by an NSERC Discovery Grant.
M.M. thanks the Research Corporation for Scientific Advancement.
The National Radio Astronomy Observatory is a facility of the National Science Foundation operated under cooperative agreement by Associated Universities, Inc.

%%%%%%%% REFERENCES %%%%%%%%%

\bibliographystyle{apj}
%\bibliography{rdf,journals1,modrefs,psrrefs,crossrefs}

\end{document}

%% file: dns_table_arXiv.tex
\begin{deluxetable*}{lrlrrrlll}
\tablecolumns{9}
\tablecaption{Known Double Neutron Star Binary Systems with Measured Component Masses.\label{tab:dns}}
\tablewidth{0pc}
\tablehead{
\colhead{Pulsar}          &  \colhead{$P_{\textrm{spin}}$}  &   \colhead{$e$}      &  \colhead{$P_{\textrm{orb}}$}  &   \colhead{$m_1$\tablenotemark{a}} &   \colhead{$m_2$\tablenotemark{a}}  &  \colhead{$V_{\textrm{tr}}$}  &  \colhead{$P_{\textrm{prec}}$\tablenotemark{b}}    &  \colhead{References} \\
\colhead{}          &  \colhead{(ms)}  &   \colhead{}      &  \colhead{(days)}  &   \colhead{($\msun$)} &   \colhead{($\msun$)}  &  \colhead{(km\,s$^{-1}$)}  &  \colhead{(years)}    &  \colhead{} 
}
\startdata 
% data
%\cline{2-8}
  J0737$-$3039A/B\tablenotemark{c}  &   22.70/2773                &   0.0878   &  0.102          &  1.3381(7)\phn                      &   1.2489(7)                         &  \phn10\tablenotemark{d}   &  \phn75   &   1   \\
  J1518+4904                        &   40.94\phn\phn\phm{/}      &   0.249    &  8.63\phn       &  $< 1.17$\phn\phn\phn\phn\phm{()}   &   $>1.55$\phn\phn\phn\phn\phm{()}   &  \phn25                    &  \nodata  &   2   \\
  B1534+12                          &   37.90\phn\phn\phm{/}      &   0.274    &  0.421          &  1.3332(10)                         &   1.3452(10)                        &  107                       &  706      &   3 \\
  J1756$-$2251                      &   28.46\phn\phn\phm{/}      &   0.181    &  0.320          &  1.312(17)\phn                      &   1.258(18)\phn                     &  \nodata\tablenotemark{e}  &  488      &   4   \\
  J1811-1736                        &   104.2\phn\phn\phn\phm{/}  &   0.828    &  18.8\phn\phn   &  $< 1.64$\phn\phn\phn\phn\phm{()}   &   $> 0.93$\phn\phn\phn\phn\phm{()}  &  \nodata                   &  \nodata  &   5,6 \\
  J1829+2456                        &   41.01\phn\phn\phm{/}      &   0.139    &  1.18\phn       &  $< 1.34$\phn\phn\phn\phn\phm{()}   &   $> 1.26$\phn\phn\phn\phn\phm{()}  &  \nodata                   &  \nodata  &   7,8 \\
  J1906+0746\tablenotemark{f}       &   144.1\phn\phn\phn\phm{/}  &   0.0853   &  0.166          &  1.290(11)\phn                      &   1.323(11)\phn                     &  \nodata                   &  169      &   9,10 \\
  B1913+16                          &   59.03\phn\phn\phm{/}      &   0.617    &  0.323          &  1.4408(3)\phn                      &   1.3873(3)\phn                     &  \phn88                    &  296      &   11   \\
  B2127+11C                         &   30.53\phn\phn\phm{/}      &   0.681    &  0.335          &  1.358(10)\phn                      &   1.354(10)\phn                     &  168                       &  278      &   12   
\enddata
\tablenotetext{a}{Here, $m_1$ and $m_2$ refer to the first-formed (usually recycled) and second-formed (usually unrecycled) neutron stars, respectively.  In all cases, the first-formed NS is observed as a pulsar, except in the case of PSR J0737$-$3039A/B, where both have been seen as pulsars.}
\tablenotetext{b}{Precession period of the first-formed neutron star in the system.}
\tablenotetext{c}{For the double pulsar we show values for both pulsars where they differ.}
\tablenotetext{d}{This value is based on a distance of 500 pc to the pulsar, based on the dispersion measure found from timing analysis.}
\tablenotetext{e}{So far only proper motion in the RA direction has been measured to be $-0.7(2)$ mas\,yr$^{-1}$, corresponding to a velocity in the RA direction of $\sim 8.3\kms$.}
\tablenotetext{f}{It is now thought that the companion to PSR~J1906+0746 is likely to be a high-mass WD, however, a NS companion is not ruled out.}
%\tablenotetext{b}{$T_{\mathrm{sys}}$ values given for the Parkes telescope are for the Multibeam (centre beam) and H-OH receivers, respectively.}\\
\tablerefs{(1)~\citet{ksm+06}; (2)~\citet{jsk+08}; (3)~\citet{sttw02}; (4)~\citet{fer08}; (5)~\citet{lcm+00}; (6)~\citet{cks+07}; (7)~\citet{clm+04}; (8)~\citet{clm+05}; (9)~\citet{lsf+06}; (10)~\citet{kas12}; (11)~\citet{wnt10}; (12)~\citet{jcj+06}.}
\end{deluxetable*}

%% file: 0737A_geometry_results_table_arXiv.tex
\begin{deluxetable*}{cllllll} 
\tablecolumns{7} 
\tablewidth{0pc} 
\tablecaption{Results from PSR J0737$-$3039A Pulse Profile Geometry Fits for $i<90\degrees$.\label{tab:geometry_result}} 
\tablehead{
\colhead{Pulse Height} & \colhead{$\alpha$} & 
\multicolumn{3}{c}{$\delta$ Upper Limits}  & 
\colhead{$\rho_1$} & \colhead{$\rho_2$} \\
\colhead{($\%$)} & \colhead{($\degrees$)} & 
\multicolumn{3}{c}{($\degrees$)}  & 
\colhead{($\degrees$)} & \colhead{($\degrees$)} \\
\colhead{} & \colhead{($\pm 68\%$)} & 
\colhead{$68\%$}  & \colhead{$95\%$}  & \colhead{$99\%$}  & 
\colhead{($\pm 68\%$)} & \colhead{($\pm 68\%$)} \\
\hline 
\multicolumn{7}{c}{One-pole emission}
}
\startdata
% One pole
30   &    $90.3_{-11.3}^{+11.5}$   &  3.9  &  26  &  50   &   $115.8_{-1.7}^{+0.4}$    &  \\
35   &    $90.2_{-18.2}^{+17.1}$   &  8.4  &  55  &  62   &   $115.0_{-2.4}^{+0.5}$    &  \\
40   &    $90.4_{-16.7}^{+16.8}$   &  6.5  &  38  &  59   &   $114.1_{-2.7}^{+0.7}$    &  \\
45   &    $90.4_{-18.2}^{+18.2}$   &  6.4  &  33  &  63   &   $113.0_{-2.7}^{+0.9}$    &  \\
50   &    $90.3_{-16.3}^{+16.2}$   &  7.2  &  40  &  68   &   $111.9_{-2.3}^{+0.7}$    &  \\
\cutinhead{Two-pole emission}
% Two poles
30   &    $90.2_{-11.4}^{+11.3}$   &  0.54  &   3.9  &   5.6    &    $18.5_{-0.42}^{+4.3}$   &  $12.1_{-0.9}^{+6.1}$  \\
35   &    $90.3_{-8.5}^{+9.1}$     &  2.3  &   5.0  &   5.9    &    $17.8_{-0.37}^{+3.5}$   &  $10.9_{-0.5}^{+5.5}$  \\
40   &    $90.2_{-16.2}^{+16.3}$   &  0.56  &   3.5  &   5.1    &    $17.8_{-0.97}^{+5.8}$   &  $11.1_{-1.3}^{+8.0}$  \\
45   &    $90.2_{-11.9}^{+11.8}$   &  0.43  &   1.7  &   3.1    &    $17.0_{-0.87}^{+4.9}$   &  $10.0_{-0.9}^{+7.3}$  \\
50   &    $90.2_{-10.9}^{+10.8}$   &  0.47  &   1.9  &   3.6    &    $16.5_{-0.44}^{+4.7}$   &   \phn$8.8_{-1.0}^{+7.4}$  
\enddata
\end{deluxetable*}